\documentclass[11pt,epsf,letterpaper]{article}%
\usepackage{setspace}
\usepackage{color}
\usepackage{amsmath}
\usepackage{amsfonts}
\usepackage{verbatim}
\usepackage{amssymb}
\usepackage{graphicx}%
\providecommand{\U}[1]{\protect\rule{.1in}{.1in}}
\begin{document}

\title{Gribov ambiguity in asymptotically AdS three-dimensional gravity}
\author{Andr\'{e}s Anabal\'{o}n$^{1}$, Fabrizio Canfora$^{2}$, Alex Giacomini$^{3}$
and Julio Oliva$^{3}$\\$^{1}$\textit{Departamento de Ciencias, Facultad de Artes Liberales, Facultad
de}\\\textit{Ingenier\'{\i}a y Ciencias, Universidad Adolfo Ib\'{a}\~{n}ez,
Vi\~{n}a Del Mar, Chile.}\\$^{2}$\textit{Centro de Estudios Cient\'{\i}ficos (CECS), Casilla 1469,
Valdivia, Chile.}\\$^{3}$\textit{Instituto de F\'{\i}sica, Facultad de Ciencias, Universidad
Austral de Chile, Valdivia, Chile.}\\{\small andres.anabalon@uai.cl, canfora@cecs.cl,
alexgiacomini@uach.cl, julio.oliva@docentes.uach.cl}}
\maketitle

\begin{abstract}
In this paper the zero modes of the de Donder gauge Faddeev-Popov operator for
three dimensional gravity with negative cosmological constant are analyzed. It
is found that the AdS$_{3}$ vacuum produces (infinitely many) normalizable
smooth zero modes of the Faddeev-Popov operator. On the other hand, it is
found that the BTZ black hole (including the zero mass black hole) does not
generate zero modes. This differs from the usual Gribov problem in QCD where
close to the maximally symmetric vacuum, the Faddeev-Popov determinant is
positive definite while \textquotedblleft far enough" from the vacuum it can
vanish. This suggests that the zero mass BTZ black hole could be a suitable
ground state of three dimensional gravity with negative cosmological constant.
Due to the kinematic origin of this result, it also applies for other
covariant gravity theories in three dimensions with AdS$_{3}$ as maximally
symmetric solution, such as new massive gravity and topologically massive
gravity. The relevance of these results for SUSY breaking is pointed out.

\end{abstract}

\section{Introduction}

The Yang-Mills interaction is paramount in the current understanding of the
fundamental interactions. As is well known, the Yang-Mills Lagrangian takes
the form
\begin{equation}
L=trF_{\mu\nu}F^{\mu\nu}\;,\;\;\;(F_{\mu\nu})^{a}=(\partial_{\mu}A_{\nu
}-\partial_{\nu}A_{\mu}+[A_{\mu},A_{\nu}])^{a},
\end{equation}
where the degrees of freedom of the theory are redundantly described by a Lie
algebra valued one form, namely the gauge potential $(A_{\mu})^{a}$. The
redundancies arise due to the invariance of the Lagrangian under finite gauge
transformations, which acts on the gauge potential as%
\begin{equation}
A_{\mu}\rightarrow U^{\dagger}A_{\mu}U+U^{\dagger}\partial_{\mu}%
U.\label{gaugetranformation}%
\end{equation}
It immediately follows that any physical observable must be invariant under
gauge transformations. As a dynamical system, the redundancy implies that the
symplectic form has a fixed, non-maximal rank, being it invertible only on the
surface where the constraints holds and the restriction of the symplectic form
to that surface is achieved by fixing the gauge. Indeed, the gauge-fixing is
quite relevant in the classical theory since, when using the Dirac bracket
formalism, the Faddeev-Popov determinant appears in the denominators of the
Dirac-Poisson brackets (see, for instance, the detailed analysis in \cite{HRT}).

It could be possible, in principle, to formulate from the very beginning the
theory in terms of gauge invariant variables such as Wilson loops. However, up
to now, such an ambitious program has been carried out completely only in the
cases of topological field theories in 2+1 dimensions \cite{wittenjones},
while it is still far from clear how to perform practical computations, such
as scattering amplitudes and correlation functions using the Wilson loop
variables for Yang-Mills theories in 2+1 and 3+1 dimensions.

Therefore, the gauge fixing seems to be unavoidable to properly describe the
evolution of a gauge theory. In Yang-Mills theories, the most convenient
choices are the Coulomb gauge and the Lorenz gauge:%
\begin{equation}
\partial^{i}A_{i}=0,\;\;\;\partial^{\mu}A_{\mu}=0\; \label{coulomblorenz}%
\end{equation}
where $i=1,..,D$ are the spacelike indices while $\mu=0,1,...,D$ are
space-time indices. Indeed, other choices are possible such as the axial
gauge, the temporal gauge, etcetera. Nevertheless these gauge fixings have
their own problems (see, for instance, \cite{DeW03}).

If it exist a proper gauge transformation\footnote{A proper gauge
transformation has to be everywhere smooth and it has to decrease fast enough
at infinity such that, a suitable norm to be specified later, converges. The
problem of defining a proper gauge transformation was first explored in
\cite{Benguria:1976in}.} (\ref{gaugetranformation}) that preserves any of the
gauge conditions (\ref{coulomblorenz}), then it would not fix the gauge
freedom completely, making impossible the avoidance of some kind of
overcounting in the gauge fixed path integral \cite{Gri78}: this phenomenon,
called \textit{Gribov ambiguity,} prevents one from obtaining a \textit{proper
gauge fixing}. Furthermore, it has been shown by Singer \cite{singer}, that if
Gribov ambiguities occur in Coulomb gauge, they occur in any gauge fixing
involving derivatives of the gauge field. Abelian gauge theories on flat
space-time, are devoid of this problem, since the Gribov copy equation for the
smooth gauge parameter $\phi$ is
\begin{equation}
\partial_{i}\partial^{i}\phi=0\;\;\;\mathrm{or}\;\;\;\partial_{\mu}%
\partial^{\mu}\phi=0 \label{abelian}%
\end{equation}
which on flat space-times (once the time coordinate has been Wick-rotated:
$t\rightarrow i\tau$) has no smooth non-trivial solutions fulfilling the
physical boundary conditions.

The situation changes dramatically when we consider an Abelian gauge field
propagating on a curved background due to the replacement of the partial
derivatives by covariant derivatives: it was shown in \cite{CGO} that, quite
generically, a proper gauge fixing in the Abelian case cannot be achieved.

In the case of non-Abelian gauge theories, even on flat space-time the Lorenz
or Coulomb gauge fixing are ambiguous. In the path integral formalism, an
ambiguity in the gauge fixing corresponds to a smooth normalizable zero mode
of the Faddeev-Popov (\textbf{FP}) determinant. In order to define the path
integral in the presence of Gribov copies, it has been suggested to exclude
classical $A_{\mu}$ backgrounds which generate zero modes of the FP operator
(see, in particular, \cite{Gri78} \cite{Zw82}-\cite{Va92}). This possibility
is consistent with the usual perturbative point of view since, in the case of
$SU(N)$ Yang-Mills theories, for a \textquotedblleft small
enough\textquotedblright\ potential $A_{\mu}$ (with respect to a suitable
functional norm \cite{Va92}), there are no zero modes of the Faddeev-Popov
operator in the Landau gauge.

It is therefore very important to analyze the problem of gauge fixing
ambiguities in the context of gravitational field. It is quite well known that
Gribov ambiguities are also present in the gravitational case (see the review
\cite{EPZ04} \cite{SS05}), but very few explicit cases have been considered.
As it will be shown below, the gravitational gauge fixing problem is quite
peculiar and very interesting from many points of view.

A very interesting case which allows an explicit analysis showing, at the same
time, many peculiar features is gravity in three dimensions \cite{Deser1y2}. Since the work by
Witten \cite{Witten:1988hc}, the standard lore has been that the
Einstein-Hilbert action defines a quantum theory of gravity in three
dimensions. In three dimensions, the traceless part of the Riemann tensor
identically vanishes and all the geometrical information is thus encoded in
the Ricci tensor. This implies that an Einstein space is locally of constant
curvature. As it was pointed out in \cite{Witten:1988hc} the theory under
consideration is thus trivial modulo the existence of global obstructions. The
seemingly uninteresting situation got closer to what one would expect from a
realistic gravitational theory with the discovery that, when the cosmological
constant is negative, one of the possible global obstructions is actually a
black hole \cite{BTZ12}.

Moreover, it is possible to give a microscopic description of the entropy of
the BTZ black hole which has its roots in the well known work of Brown and
Henneaux \cite{Brown:1986nw}: they showed that the asymptotic symmetry group
of a three dimensional spacetime which matches $AdS_{3}$ at infinity (with a
precise fall off) is the product of two copies of the de Witt algebra and that
its canonical realization is projective. The existence of such central charge
was connected, much later, with the entropy of this black hole through the
Cardy formula \cite{cardy, strom}. The fact that this result holds even for
hairy black holes \cite{Correa:2010hf} suggests that such a result is
independent on the finiteness of three dimensional quantum gravity. Indeed,
since pure gravity is classically trivial any possible counter-term is
on-shell equivalent to the volume of the spacetime and can be reabsorbed in
the cosmological constant and a local redefinition of the metric tensor.
Furthermore, the fact that 3-D gravity can be formulated as a Chern-Simons
gauge theory implies that the cosmological constant is proportional to the
structure functions of the gauge group, and the non existence of gauge
anomalies in three dimensions can be used to argue that the gauge algebra
holds at the quantum level. Therefore, no renormalization can affect the
cosmological constant and the theory must be finite \cite{Witten:2007kt}, this
argument support the Strominger proposal that three dimensional Einstein
gravity \emph{is} a conformal field theory. However, when there are matter
fields, the first ingredient of the previous discussion no longer holds,
namely the fact that on-shell every spacetime has constant curvature. Although
it cannot be discarded that there is an ultraviolet completion of the theory
analyzed in \cite{Correa:2010hf}, the Cardy formula gives the right result
without the arguments that Strominger propose to use it.

Due to the triviality of the theory one would expect to obtain the
thermodynamical properties of the BTZ black hole computing the partition
function \cite{maloneywitten}, however the sum of all the contributions coming
from known classical geometries gives rise to an inconsistent result.

Our aim in writing this paper is to point out that the quantization process
should be reanalyzed at the light of the existence of an infinite number of
zero modes of the FP operator for the de Donder gauge in three dimensions when
gravity is quantized around $AdS_{3}$ while it has no non-trivial zero mode
for the BTZ black holes.

The paper is organized as follows: section 2 presents the generalities of the
Gribov problem associated to diffeomorphism invariance in a gravitational
theory. In section 3 we focus in the case of $AdS_{3}$, where we found a set
of vector fields which generate Gribov copies for the de Donder Faddeev-Popov
operator, which preserve the asymptotically $AdS_{3}$ behavior of the metric,
in the Brown-Henneaux sense. Section 4 is devoted to analyze the same problem
on BTZ black holes and we prove that, within the family of diffeomorphisms
considered here, there are no normalizable copies, (suggesting then massless
BTZ black hole as a better ground-state to perform a perturbative analysis).
Finally in section 5 we give some further comments concerning to other
theories of gravity and a possible new approach for supersymmetry breaking.

\section{\bigskip Gribov ambiguity in gravitational theories}

The degrees of freedom of the gravitational field are described by the metric
tensor $g_{\mu\nu}$. Due to the diffeomorphism invariance, metric tensors
related by a coordinate transformation describe the same space-time. In the
framework of the path integral approach to the semi-classical quantization of
gravity, the most commonly used gauge fixing condition of the diffeomorphism
invariance is the de Donder gauge defined as follows\footnote{It is worth
noting that this gauge is also the most common choice in the analysis of
gravitational waves \cite{MTW}.}. Let us denote the classical background
metric as $g_{\mu\nu}^{(0)}$, while a small fluctuation around $g_{\mu\nu
}^{(0)}$ will be denoted as $h_{\mu\nu}$. Under a generic coordinate
transformation, the metric fluctuation $h_{\mu\nu}$ transforms as%
\begin{equation}
h_{\mu\nu}\rightarrow h_{\mu\nu}+\nabla_{\mu}\xi_{\nu}+\nabla_{\nu}\xi_{\mu}
\label{gatra1}%
\end{equation}
where $\xi_{\mu}$\ is the vector field generating the diffeomorphism and the
covariant derivative is taken with the background metric $g_{\mu\nu}^{(0)}$.
Then, the de Donder gauge on the metric fluctuation $h_{\mu\nu}$ reads%
\begin{equation}
\nabla^{\mu}h_{\mu\nu}=0\ , \label{dedo1}%
\end{equation}
The field equations of general relativity in vacuum further imply that%
\begin{equation}
h_{\ \mu}^{\mu}=0\ ; \label{dedo2}%
\end{equation}
on the other hand, if one is analyzing a diffeomorphism invariant theory
different from general relativity, only the condition (\ref{dedo1}) should be
taken into account. In the present paper, we will mainly focus on general
relativity and some comments on alternative theories of gravity will be
presented in the last section.

In the context of three-dimensional quantum general relativity, the analysis
of gauge-fixing problem is even more relevant than in four-dimensional
space-times because the theory is finite \cite{Witten:1988hc}. Indeed, in this
case it is natural to use the Lorentzian path integral. The reason to perform
the Wick rotation in Yang-Mills, is due to the fact that in this case the
Euclidean action is bounded from below. Also in asymptotically flat spacetimes
in four dimensions, the Euclidean Einstein-Hilbert action is positive definite
\cite{S-Y}. Nevertheless, in three dimensional asymptotically AdS gravity,
this is not the case anymore, and then there is not mandatory reason to
consider the Euclidean path integral. Furthermore, in the case of gravity in
2+1 dimensions, it is also possible to consider explicitly topology change
amplitudes \cite{witten2} using the Lorentzian path integral. The interest in
considering the issue of topology change amplitudes is based on the
spin-statistics connection, which strongly suggests the necessity of topology
changes in quantum gravity \cite{sorkin1} \cite{sorkin2} (for detailed reviews
see \cite{bala} \cite{bala1}). For these reasons, we will consider in the
present paper the Lorentzian signature.

\bigskip

The main goal of this paper is to show that the gauge fixing ambiguity for
three-dimensional general relativity with negative cosmological constant and,
more generically, for any diffeomorphism invariant metric theory is actually
quite different from the usual Gribov problem for $SU(N)$ Yang-Mills in a four
dimensional flat space-time. The maximally supersymmetric vacuum of the theory
(which is AdS$_{3}$ space-time) generates a denumerable set of, normalizable,
smooth, zero modes of the Faddeev-Popov (FP) operator in the de Donder gauge.
Thus, even if following the QCD case in four dimensions one would naturally
expect \cite{witten2} that the Gribov problem should not manifest itself at a
perturbative level, even perturbative calculations in the path integral
formalism around the maximally supersymmetric vacuum seems to be problematic.
On the other hand, it is found that the BTZ black hole solutions of this
theory \cite{BTZ12} do not generate zero modes. In particular, the massless
BTZ black hole, appears to be a sensible ground state for perturbative
analysis but \textit{it preserves only half of the supersymmetries}

It is also worth to emphasize that even if one would find another gauge fixing
free of Gribov ambiguities, still the presence of Gribov ambiguities in the De
Donder gauge would have very deep physical consequences. In particular, it is
known that the presence of Gribov ambiguity can led to a breaking of the BRST
symmetry at a non-perturbative level (see, for instance, \cite{Fuj}%
-\cite{BaSo}). Therefore, even if one would adopt an ambiguity-free gauge
fixing, it would be still necessary to analyze carefully the above issues.

\section{Zero modes of the FP operator on $AdS_{3}$ background}

The metric of AdS$_{3}$ is%
\begin{equation}
ds^{2}=-\left(  1+\frac{r^{2}}{l^{2}}\right)  dt^{2}+\frac{dr^{2}}%
{1+\frac{r^{2}}{l^{2}}}+r^{2}d\phi^{2}\ ; \label{ads1}%
\end{equation}
where
\begin{equation}
-\infty<t<\infty,\qquad0\leq r<\infty,\qquad0\leq\phi<2\pi.
\end{equation}

The two copies of the de Witt algebra that arise as a non-trivial endomorphism
of the space of solutions of general relativity with negative cosmological
constant $\Lambda=-1/l^{2}$ in three dimensions, has a description in terms of
the asymptotic Killing vectors which preserve the asymptotic behavior of the
metric at spatial infinity \cite{Brown:1986nw}. The asymptotic Killing vectors
read%
\begin{align}
\xi^{t}  &  =l\left(  T^{+}+\frac{l^{2}}{2r^{2}}\partial_{-}^{2}T^{-}%
+T^{-}+\frac{l^{2}}{2r^{2}}\partial_{+}^{2}T^{+}\right)  +O\left(
r^{-4}\right) \label{bh1}\\
\xi^{\phi}  &  =T^{+}-T^{-}+\frac{l^{2}}{2r^{2}}\partial_{-}^{2}T^{-}%
-\frac{l^{2}}{2r^{2}}\partial_{+}^{2}T^{+}+O\left(  r^{-4}\right)
\label{bh2}\\
\xi^{r}  &  =-r\left(  \partial_{+}T^{+}+\partial_{-}T^{-}\right)  +O\left(
r^{-1}\right)  \ . \label{bh3}%
\end{align}
where $x^{\pm}=\frac{t}{l}\pm\phi$ ($\partial_{\pm}=\frac{l}{2}\partial_{t}%
\pm\frac{1}{2}\partial_{\phi}$), and
\begin{equation}
T^{\pm}:=T^{\pm}\left(  \frac{t}{l}\pm\phi\right)  ,
\end{equation}
which preserve the following asymptotic behavior of the metric%
\begin{equation}
h_{rr}\sim O\left(  r^{-4}\right)  ,\ h_{rm}\sim O\left(  r^{-3}\right)
,\ h_{mn}\sim O\left(  1\right)  , \label{asympBH}%
\end{equation}
where the indices $m,n$ stand for $\left\{  t,\phi\right\}  $, and $h_{\mu\nu
}$ is the departure from the $AdS_{3}$ spacetime.\bigskip

Obviously, because of the periodicity of the coordinate $\phi$ both the
coordinate $x^{+}=\frac{t}{l}+\phi$ and the coordinate $x^{-}=\frac{t}{l}%
-\phi$ are periodic:
\begin{equation}
x^{\pm}\sim x^{\pm}+2\pi.\label{percor}%
\end{equation}
Therefore, it is possible to Fourier analyze the functions $T^{\pm}\left(
\frac{t}{l}\pm\phi\right)  $. These modes furnish a realization of two copies
of the de Witt algebra with the Lie bracket. On the other hand, if the
periodicity in the coordinate $\phi$ is disregarded the arbitrary functions in
the Killing vectors can be expanded in Laurent series, and the same asymptotic
algebra can be obtained \cite{Barnich:2010eb}. In the forthcoming calculations
we will adopt the expansion in Fourier modes.

In order to construct zero modes of the FP operator in the de Donder gauge on
the AdS$_{3}$ background metric (\ref{ads1}), one has to solve the following
system of equations%
\begin{equation}
\nabla^{\mu}\left(  \nabla_{\mu}\eta_{\nu}+\nabla_{\nu}\eta_{\mu}\right)  =0
\label{zm1}%
\end{equation}
together with the scalar equation%
\begin{equation}
\nabla^{\mu}\eta_{\mu}=0\ , \label{zm2}%
\end{equation}
where $\nabla_{\mu}$\ is the covariant derivative with respect to the
$AdS_{3}$ metric in equation (\ref{ads1}). A proper, linearized
diffeomorphism, $\eta_{\mu}$, has to be smooth everywhere, and furthermore it
has to posses a finite norm $\mathcal{N}(\eta)$:%
\begin{equation}
\mathcal{N}(\eta):=\int\sqrt{-g}d^{3}x\ \nabla_{\left(  \mu\right.  }%
\eta_{\left.  \nu\right)  }\nabla^{\left(  \mu\right.  }\eta^{\left.
\nu\right)  }<\infty\ . \label{norm1}%
\end{equation}
It is worth noting here that the above norm in equation (\ref{norm1}) is the
closest analogue of the functional norm used in the Yang-Mills path integral
(see in particular \cite{Va92}). In the case of the Yang-Mills path integral,
the gauge potential\footnote{In the case of Yang-Mills path integral, the
gauge potential $A_{\mu}^{a}$ is "a small fluctuation". Namely, $A_{\mu}^{a}$
represents a small deviation from the maximally symmetric vacuum $A_{\mu}%
^{a}=0$ (or any other classical background one is interested in). The norm
conditions are supposed to describe mathematically the "smallness" of the
fluctuations.} $A_{\mu}^{a}$ has to satisfy the following finite norm
condition%
\begin{equation}
\mathcal{N}_{YM}(A)=\int\sqrt{-g}d^{D+1}x\ Tr\left(  A_{\mu}A^{\mu}\right)
<\infty\ . \label{fn1}%
\end{equation}
As it is well known, this induces a condition on the gauge transformation
parameter $U$ in Eq. (\ref{gaugetranformation}) by requiring that whenever
$A_{\mu}^{a}$ satisfies the condition in Eq. (\ref{fn1}) also the gauge
transformed of $A_{\mu}^{a}$ has to satisfy it. In the gravitational case in
(2+1) dimensions, the path integral is taken on the metric fluctuation
$h_{\mu\nu}$. The natural norm for $h_{\mu\nu}$ is then%
\begin{equation}
\mathcal{N}(h):=\int\sqrt{-g}d^{3}x\ h_{\mu\nu}h^{\mu\nu}\ , \label{nogra1}%
\end{equation}
and the finite norm condition is $\mathcal{N}(h)<\infty$. This norm induces
then the condition on the vector field $\eta_{\mu}$ given in equation
(\ref{norm1}).

\bigskip

Let us consider the following ansatz for a vector field $\eta^{\mu}$ which is
designed in order to accommodate explicitly the Brown-Henneaux asymptotics in
Eqs. (\ref{bh1}), (\ref{bh2}) and (\ref{bh3})%
\begin{align}
\eta^{+}  &  =f_{1}\left(  r\right)  T^{+}+f_{2}\left(  r\right)  \frac{l^{2}%
}{2r^{2}}\partial_{-}^{2}T^{-}\ ,\label{ans1}\\
\eta^{-}  &  =f_{3}\left(  r\right)  T^{-}+f_{4}\left(  r\right)  \frac{l^{2}%
}{2r^{2}}\partial_{+}^{2}T^{+}\ ,\label{ans2}\\
\eta^{r}  &  =-\frac{r}{2}\left(  f_{5}\left(  r\right)  \partial_{+}%
T^{+}+f_{6}\left(  r\right)  \partial_{-}T^{-}\right)  \ , \label{ans3}%
\end{align}
We will search for zero modes of the Faddeev-Popov operator in the de Donder
gauge within the family determined by (\ref{ans1})-(\ref{ans3}), such that the
functions $f_{\left(  i\right)  }\left(  r\right)  $ ($i=1$, ...,$6$) are
everywhere smooth and guarantee that the norm in Eq. (\ref{norm1}) is finite.
The diffeomorphisms generated by the vector field $\eta^{\mu}$ with components
(\ref{ans1})-(\ref{ans3}) will belong to the Brown-Henneaux class
(\ref{bh1})-(\ref{bh3}), provided the functions $f_{\left(  i\right)  }(r)$
satisfy the following asymptotic behavior%
\begin{equation}
f_{\left(  i\right)  }\left(  r\right)  \rightarrow_{r\rightarrow+\infty
}\left\{
\begin{array}
[c]{ccc}%
\alpha+O\left(  r^{-4}\right)  & \text{for} & i=1,3\\
& \text{and} & \\
\beta+O\left(  r^{-2}\right)  & \text{for} & i=2,4,5,6
\end{array}
\right.  \ , \label{asympis}%
\end{equation}
where $\alpha$ and $\beta$ are constant.

It is possible to Fourier expand $T^{+}\left(  x^{+}\right)  $ and
$T^{-}\left(  x^{-}\right)  $ so that one can replace $T^{+}$ and $T^{-}$ as
follows:%
\begin{align}
T^{+}  &  \rightarrow e^{inx^{+}}\ ,\label{ansatzangular}\\
T^{-}  &  \rightarrow e^{imx^{-}}\ .
\end{align}
where the expansion in Fourier modes is in the interval when the function is
non-zero. Here we consider $n\geq2$ and $m\geq2$. The computation with
negative $n$ and $m$ follows the same lines. The modes with $m,m\in\left\{
0,\pm1\right\}  $ which generate the $sl\left(  2,R\right)  \times sl\left(
2,R\right)  $ subalgebra of the asymptotic Brown-Henneaux symmetries, will be
discussed later.

With the above requirements, and with an ansatz for $\eta^{\mu}$ of the form
(\ref{ans1})-(\ref{ans3}), the relevant solutions for the system (\ref{zm1})
and (\ref{zm2}) are%
\begin{align*}
f_{1}\left(  r\right)   &  =C_{3}\frac{r^{n-2}}{2\left(  r^{2}+l^{2}\right)
^{\frac{n+2}{2}}}\left(  4r^{4}+2l^{2}\left(  n+2\right)  r^{2}+nl^{4}\left(
n+1\right)  \right)  \ ,\\
f_{2}(r)  &  =-\frac{\left(  4l^{2}+5r^{2}\right)  r^{3}}{l^{4}m^{2}}%
f_{6}^{\prime}-\frac{\left(  l^{2}+r^{2}\right)  r^{4}}{l^{4}m^{2}}%
f_{6}^{\prime\prime}+\frac{r^{2}\left(  l^{2}m^{2}-2l^{2}-2r^{2}\right)
}{l^{2}m^{2}\left(  l^{2}+r^{2}\right)  }f_{6}\\
f_{3}\left(  r\right)   &  =C_{4}\frac{r^{m-2}}{2\left(  r^{2}+l^{2}\right)
^{\frac{m+2}{2}}}\left(  4r^{4}+2l^{2}\left(  m+2\right)  r^{2}+ml^{4}\left(
m+1\right)  \right)  \ ,\\
f_{4}(r)  &  =-\frac{\left(  4l^{2}+5r^{2}\right)  r^{3}}{l^{4}n^{2}}%
f_{5}^{\prime}-\frac{\left(  l^{2}+r^{2}\right)  r^{4}}{l^{4}n^{2}}%
f_{5}^{\prime\prime}+\frac{r^{2}\left(  l^{2}n^{2}-2l^{2}-2r^{2}\right)
}{l^{2}n^{2}\left(  l^{2}+r^{2}\right)  }f_{5}\ ,\\
f_{5}\left(  r\right)   &  =\frac{C_{3}r^{n-2}}{\left(  r^{2}+l^{2}\right)
^{\frac{n}{2}}}\left(  nl^{2}+l^{2}+2r^{2}\right)  \ ,\\
f_{6}\left(  r\right)   &  =\frac{C_{4}r^{m-2}}{\left(  r^{2}+l^{2}\right)
^{\frac{m}{2}}}\left(  l^{2}m+l^{2}+2r^{2}\right)  \ ,
\end{align*}
where we used the notation $X^{\prime}:=\partial_{r}X\ .$

The asymptotic behavior at infinity of these functions is given by%
\begin{align*}
f_{1}\left(  r\right)   &  =2C_{3}+C_{3}\frac{l^{4}n^{2}}{r^{4}}+O\left(
r^{-6}\right)  \ ,\\
f_{2}(r)  &  =2C_{4}+C_{4}\frac{l^{2}\left(  1-m-m^{2}\right)  }{mr^{2}%
}+O\left(  r^{-4}\right)  \ ,\\
f_{3}\left(  r\right)   &  =2C_{4}+C_{4}\frac{l^{4}m^{2}}{r^{4}}+O\left(
r^{-6}\right)  \ ,\\
f_{4}(r)  &  =2C_{3}+C_{3}\frac{l^{2}\left(  1-n-n^{2}\right)  }{nr^{2}%
}+O\left(  r^{-4}\right)  \ ,\\
f_{5}\left(  r\right)   &  =2C_{3}+\frac{C_{3}l^{2}}{r^{2}}+O\left(
r^{-4}\right)  \ ,\\
f_{6}\left(  r\right)   &  =2C_{4}+\frac{C_{4}l^{2}}{r^{2}}+O\left(
r^{-4}\right)  \ ,
\end{align*}
fulfilling then the required behavior given by (\ref{asympis}).

The radial part of the integral contributing to the norm in equation
(\ref{norm1}) is given by%
\begin{equation}
\mathcal{N}\varpropto\frac{1}{l^{2}\left(  m+n-2\right)  }\left.
\frac{r^{m+n-2}}{\left(  r^{2}+l^{2}\right)  ^{\frac{m+n-2}{2}}}\right\vert
_{0}^{\infty}\ .\label{NFinal}%
\end{equation}
which converges for $n,m\geq2$, since, when one considers the variables
$x^{+}$ and $x^{-}$ periodic as in Eq. (\ref{percor}), the integral in $x^{+}$
and $x^{-}$ in the norm is obviously finite. On the other hand, one could
disregard the periodicity in $x^{+}$ and $x^{-}$ and consider the dependence
of the copies on these variables to be given by a function of compact support.
A careful analysis for the case $m,n\in\left\{  0,\pm1\right\}  $, along the
same lines than the one presented here, shows that within the family we
considered, there are no normalizable vector field $\eta$ which asymptotically
matches the Brown-Henneaux vector generating the $sl(2,R)\times sl(2,R)$
subalgebra\footnote{The vector fields that belong to the $sl(2,R)\times
sl(2,R)$ subalgebra of the asymptotic symmetries, are special in the sense
that they generate all the nontrivial charges associated to the known
solutions of three dimensional general relativity.}, which would generate zero
modes for the FP determinant.

\section{BTZ black hole background}

In the previous section it has been shown that for three dimensional gravity
with a negative cosmological constant, there is a gauge fixing ambiguity for
the maximally (super)symmetric vacuum. As mentioned above, this situation is
quite different from the usual Yang-Mills $SU(N)$ Gribov problem, where "near
to the vacuum" the gauge fixing is well defined. Thus, in order to get a
perturbatively well defined theory it is sufficient to restrict the path
integral to this region. The boundary in the space of connections which
delimits the region where the gauge fixing is not ambiguous is called the
Gribov horizon. Since for the gravitational field in $\left(  2+1\right)  $
dimensions the most natural ground state manifests gauge fixing ambiguities,
one would naively expect that going "far away" from the ground state the
situation could become worse. Surprisingly enough, this is not the case.
Indeed, let us consider the BTZ black hole metric with Lorentzian signature
\begin{equation}
ds^{2}=\frac{dr^{2}}{\frac{r^{2}}{l^{2}}-\mu}-\frac{l^{2}}{4}\left(
dx^{-2}+dx^{+2}\right)  -\left(  r^{2}-\frac{\mu l^{2}}{2}\right)
dx^{+}dx^{-}\
\end{equation}
where $\mu$ is a mass parameter (the AdS$_{3}$ vacuum turns out to have
$\mu=-1$). Solutions with $-1<\mu<0$ represent naked singularities and must
therefore be discarded. The other physically sensible solutions are therefore
given by $\mu>0$, which are black holes and the "zero mass black hole" is
obtained when $\mu=0$. None of these black hole solutions preserve all the
supersymmetries, nonetheless the case $\mu=0$ is half BPS. These states are
separated from the AdS vacuum by a mass gap so that they are not connected to
it and therefore they are "far away" in the space of solutions.\newline Let us
therefore analyze the existence of zero modes of the Faddeev-Popov operator
for these other asymptotically AdS solutions. For the BTZ black holes with
mass parameter larger than zero the equation for the zero modes can again be
integrated and the functions $f_{\left(  j\right)  }$ of our ansatz
(\ref{ans1})-(\ref{ans3}) acquire now the form
\begin{equation}
f_{\left(  j\right)  }\left(  r\right)  =(r-r_{+})^{i\alpha}%
\end{equation}
$\alpha$ being a real constant and $r_{+}:=l\sqrt{\mu}$. Therefore, the
function $f_{\left(  j\right)  }$ in this case posses an essential
singularity. Consequently, they do not describe a smooth proper gauge
transformation. For the massless case the function $f_{\left(  j\right)  }$
takes the form
\begin{equation}
f_{\left(  j\right)  }\left(  r\right)  =c_{1}+\frac{c_{2}}{r^{a}}+c_{3}r^{b}%
\end{equation}
with $c_{i}$ and $a,b$ real constants, which blows up at the origin or in the
asymptotic region and so again, does not give rise to a well-defined gauge
transformation $\eta$. Therefore, within the family of vector fields
considered here, it is not possible to generate a zero mode for FP operator of
the diffeomorphism invariance on BTZ black hole not even in the massless case.
This means that in three dimensions the Gribov problem for the gravitational
field seems to be reversed, in the sense that there exist a horizon around the
natural ground state such that \textit{inside} this horizon the gauge fixing
is ambiguous whereas \textit{outside} it is not. This suggests that the
massless black hole could to be a suitable vacuum for the theory, even if it
preserves less (super)symmetries than the AdS$_{3}$ space-time.

\section{Further Comments}

\begin{itemize}
\item Gauge fixing in alternative theories
\end{itemize}

The analysis of the Gribov ambiguity in other diffeomorphism invariant
theories in (2+1)-dimensions is very similar to the one presented in the
previous sections. However, there is an important difference: unlike the
condition in Eq. (\ref{zm1}) which is common to all the diffeomorphism
invariant theories, the condition in Eq. (\ref{zm2}) is particular of general
relativity. Therefore, when one searches for zero modes of the Faddeev-Popov
operator in the de Donder gauge, the condition in Eq. (\ref{zm2}) has to be
dropped. This could be relevant, for instance, in the analysis of Chiral
Gravity \cite{strom}, where the de Donder gauge has been used. In general,
this implies that in the cases of different covariant gravity theories, the
Gribov copies would be less restricted than in general relativity.

\begin{itemize}
\item An effective mechanism of (partial) supersymmetry breaking?
\end{itemize}

The present results are quite peculiar when compared with the usual Gribov
problem in $SU(N)$ Yang-Mills theory in four dimensions. In the present case,
near the maximally supersymmetric vacuum ($AdS_{3}$) there are gauge fixing
ambiguities, while \textquotedblleft far enough" from it, and within the
family of diffeomorphisms considered here, there are no gauge fixing problems.
Therefore, the Gribov problem could be used as an effective mechanism of
partial supersymmetry breaking (at least in 2+1 dimensions) since, in order to
properly define the Faddeev-Popov operator, one should consider small
fluctuations around a ground state which preserves only one half of the
supersymmetry. In other words, the appearance of Gribov copies would select a
different ground state than the one which would be selected according to the
criterion of the maximum number of supersymmetries. This issue is even more
apparent, if one considers the common point of view (see, for instance,
\cite{Gri78} \cite{Zw82}-\cite{Va92}) to cut from the path integral the
classical backgrounds affected by the presence of copies, then the maximally
supersymmetric background (AdS$_{3}$) should be excluded. In this case, one
could have at most a classical background (the zero mass BTZ black hole)
preserving one half of the supersymmetries. This is very interesting since the
problem to find a satisfactory mechanism of supersymmetry breaking has not
been solved yet (see, for instance, \cite{susybr1} \cite{susybr2}
\cite{susybr3}).

\section{Acknowledgments}

We thank Marc Henneaux for important suggestions and encouraging comments.
This work is supported by Fondecyt grants 11080056 and 11090281, and by the
Conicyt grant \textquotedblleft Southern Theoretical Physics
Laboratory\textquotedblright\ ACT-91. XXThis work has been partially funded by
the following Fondecyt grants: 11080056, 11090281, by UACh-DID grant
S-2009-57, and by the Conicyt grant \textquotedblleft Southern Theoretical
Physics Laboratory\textquotedblright\ ACT-91. The Centro de Estudios Cient%
\'{}%
\i ficos (CECS) is funded by the Chilean Government through the Centers of
Excellence Base Financing Program of Conicyt. F. C. is also supported by
PROYECTO INSERCI\'{O}N CONICYT 79090034 and by the Agenzia Spaziale Italiana (ASI).

\end{document}